\documentclass[aps,showpacs,floatfix,preprint]{revtex4}
\usepackage{amssymb,amsmath}
\usepackage[dvips]{graphicx,color,pstricks}
\usepackage{subfigure}

\begin{document}

\title{Chiral condensate  and chemical freeze-out} 
\author{D.~B.~Blaschke}
\email{blaschke@ift.uni.wroc.pl}
\affiliation{Institute for Theoretical Physics, University of Wroclaw,
50-204 Wroclaw, Poland}
\affiliation{Bogoliubov  Laboratory of Theoretical Physics, JINR,
141980  Dubna, Russia}
\author{J.~Berdermann}
\affiliation{DESY Zeuthen, 
D-15738 Zeuthen, Germany}
\author{J.~Cleymans}
\affiliation{UCT-CERN Research Centre and Department of Physics,
Rondebosch 7701, Cape Town, South Africa}
\author{K.~Redlich}
\affiliation{Institute for Theoretical Physics, University of Wroclaw,
50-204 Wroclaw, Poland}
\affiliation{ExtreMe Matter Institute EMMI, GSI, D-64291 Darmstadt, Germany}

\date{\today}

\begin{abstract}
We consider a chemical freeze-out mechanism which is based on a strong
medium dependence of the rates for inelastic flavor-equilibrating collisions
based on the delocalization of hadronic wave functions and growing
hadronic radii when approaching the chiral restoration.
We investigate the role of mesonic (pion) and baryonic (nucleon) fluctuations
for melting the chiral condensate in the phase diagram in the ($T,\mu$)-plane.
We apply the PNJL model beyond mean-field and present an effective
generalization of the chiral perturbation theory result which accounts for the 
medium dependence of the pion decay constant while preserving the GMOR relation.
We demonstrate within a schematic resonance gas model consisting of a variable
number of pionic and nucleonic degrees of freedom that within the above model
a quantitative explanation of the hadonic freeze-out curve
and its phenomenological conditions  can be given.
\end{abstract}


\maketitle


\setcounter{footnote}{0}

\section{Introduction}

The investigation of the phase diagram of QCD in the plane of temperature $T$
and baryochemical potential $\mu_B$ is one of the goals {\bf of} heavy-ion
collision experiments, of lattice QCD (LQCD) and effective field theory
approaches to the nonperturbative
{sector of QCD.}
Of particular interest are the conditions under which the
{approximate}
chiral symmetry of the QCD Lagrangian will get restored and whether this
transition will be necessarily accompanied by the deconfinement of quarks and
gluons.
More detailed questions to the QCD phase diagram concern the order of these 
phase transitions, their critical exponents and fluctuation measures as well
as the possible existence of a critical point or even a triple point in the
{phase diagram}.
Most promising tools for the experimental determination of these
characteristics are the energy scan programs at CERN, RHIC and at the upcoming
dedicated facilities of the third generation: FAIR and NICA.
The systematic analysis of higher moments of distributions of produced
particles in their dependence on the collision energy and the size of collising
systems shall provide answers to the above questions and allow direct
comparison with predictions from the underlying theory, as provided by LQCD,
 see, e.g., Ref.~\cite{Karsch:2010ck} and works cited therein.

As long as the applicability of LQCD methods is bound to the region of
finite $T$ {and}  {$\mu_B/T \ll 1$}, any predictions for the
phase structure of QCD at high baryon densities including the possible
existence of critical points will rely on  effective {models}.
To  be relevant for the discussion of the above problems these {models} have
to share with QCD the property of chiral symmetry and its dynamical breaking
as well as a mechanism for confinement and deconfinement.

{At  { present} stage, of particular  relevance for the discussion of the
QCD phase diagram are the chemical freeze-out parameters ($T^{f},\mu_B^{f}$)
which have been obtained from the  statistical model analysis of  particle
yields obtained in heavy ion collisions
\cite{BraunMunzinger:2003zd,BraunMunzinger:2001ip}.}
One of the most striking observations is the systematic behaviour of these
parameters with collision energy $\sqrt{s}$
\cite{Cleymans:1998fq,Cleymans:1999st,Cleymans:2005xv},
which has recently been given a simple parametric form \cite{Cleymans:2006qe}.
It has been observed that the {resulting} freeze-out curve in the phase diagram
is closely correlated to the thermodynamical quantities of the hadron
resonance gas described by the statistical model.
These phenomenological freeze-out {conditions}  make statements about the mean
energy per hadron $\langle E \rangle / \langle N \rangle \simeq 1.0$ GeV,
the dimensionless {entropy density $s/T^3\simeq 7$ and a total baryon and
antibaryon density $n_B + n_{\bar{B}} \simeq  0.12$ fm$^{-3}$.}
The freeze-out line provides a lower bound for the chiral restoration and
deconfinement {transition}  in the phase diagram.
Being coincident at low densities as inferred from LQCD \cite{Karsch:1994hm},
both transitions need not to occur simultaneously at high density thus allowing
for an island of a quarkyonic phase \cite{McLerran:2007qj}
between hadron gas and quark-gluon plasma
with a (pseudo-)triple point \cite{Andronic:2009gj}.

The question appears for the physical mechanism which governs the chemical
freeze-out and which determines quantitatively the freeze-out parameters.
One aspect is provided by the requirement that hadrons should overlap in order
to facilitate flavor exchange reactions which establish chemical equilibrium.
This geometrical picture of freeze-out is sucessfully realized in a percolation
theory approach \cite{Magas:2003wi}.
Another aspect is the dynamical one: when the equation of state (EoS) possess
softest points (e.g., due to the dissociation of hadrons into their quark and
gluon constituents with a sufficient release of binding energy involved) which
is correlated with the freeze-out curve in the phase diagram, then it is
obvious that the hadron abundances are characterized by the corresponding $T$
and $\mu_B$ values \cite{Toneev:2000ym}.
The dynamical system got quasi trapped at the softest points for sufficient
time to achieve chemical equilibration before evaporating as a gas of hadron
resonances freely streaming to the particle detectors. 
{Also the kinetic aspect of fast  chemical  equilibration was discussed in the 
hadronic gas when accounting for a multi-hadron dynamics \cite{pbm}.}

{All these} mechanisms are appealing since they provide an intuitively clear picture,
but they are flawed by the fact that their relation to fundamental aspects of
the QCD phase transition like the chiral condensate as an order parameter are
not an element of the description.

In the present contribution we develop an approach which relates the
geometrical as well as the hydrodynamic and kinetic aspects of chemical
freeze-out to the medium dependence of the chiral condensate.
We demonstrate within a beyond-meanfield extension
\cite{Schaefer:2007pw,Blaschke:2007np,Rossner:2007ik,Blaschke:2009qk,Skokov:2010wb,Skokov:2010uh,Radzhabov:2010dd}
of the Polyakov NJL model
\cite{Fukushima:2003ib,Fukushima:2003fw,Fukushima:2003fm,Fukushima:2002ew,Ratti:2005jh,Roessner:2006xn,Sasaki:2006ww}
how the excitation of hadronic resonances initiates the melting
of the chiral condensate which entails a Mott-Anderson type delocalization of
the hadron wave functions and a sudden drop in the relaxation time for flavor
equilibration.
Already for a schematic resonance gas consiting of pions and nucleons with
artificially enhaced numbers of degrees of freedom we can demonstrate that a
kinetic freeze-out condition for the above model provides quantitative
agreement with the phenomenological freeze-out curve.

\section{Chemical freeze-out and chiral condensate}

We propose
to  relate the chemical freeze-out to the chiral condensate in the following
effective way.
As a freeze-out condition for flavor equilibrating
reaction kinetics in the temperature-chemical potential plane
$(T,\mu)$ we employ

\begin{equation}
\tau_{\rm exp}(T,\mu)=\tau_{\rm coll}(T,\mu)~,
\label{fo}
\end{equation}
where $\tau_{\rm exp}(T,\mu)$ is the expansion time scale of the hadronic
fireball and the inverse of the relaxation time for reactive collisions is
\begin{equation}
\label{freezeout}
\tau_{\rm coll}^{-1}(T,\mu)=\sum_{i,j}\sigma_{ij}n_j~,
\end{equation}
with $i,j=\pi, N, ...$ running over all species in the hadron resonance gas. 
For the cross sections we adopt the geometrical Povh-H\"ufner law
\cite{Hufner:1992cu,Povh:1990ad}
\begin{equation}
\label{ph}
\sigma_{ij}=\lambda \langle r_i^2 \rangle \langle r_j^2 \rangle~,
\end{equation}
where $\lambda$ is a constant of the order of the string tension
$\lambda\sim 1~{\rm GeV/fm}= 5~{\rm fm}^{-2}$.
Note that this behaviour has been obtained for the quark exchange
contribution to hadron-hadron cross sections \cite{Martins:1994hd}.

A key point of our approach is that the radii of hadrons shall depend on
$T$ and $\mu$ and shall diverge when hadron dissociation (Mott effect) sets
in, driven basically by the restoration of chiral symmetry.
This has quantitatively been studied for the pion \cite{Hippe:1995hu}, where
it has been shown that close to the Mott transition the chiral perturbation
theory corrections can be safely neglected and the pion radius is well
approximated by

\begin{equation}
r_\pi^2(T,\mu)=\frac{3}{4\pi^2} f_\pi^{-2}(T,\mu)~.
\end{equation}

It has been demonstrated that the GMOR relation holds out to the chiral
phase transition where pions would merge the continuum of unbound quark matter
\cite{Blaschke:1999ab,Blaschke:2000gd}.
Since the current quark mass is  $T-$ and $\mu -$independent and the
pion mass is ``chirally protected'',
the  $T-$, $\mu-$dependence of the chiral condensate has to be reflected in a
similar behaviour of the pion decay constant
\begin{equation}
\label{GMOR}
f_\pi^2(T,\mu)=-m_0\langle \bar{q}q\rangle_{T,\mu}/M_\pi^2.
\end{equation}
The resulting relationship between pion radius and chiral condensate in the
medium reads
\begin{equation}
r_\pi^2(T,\mu)=\frac{3M_\pi^2}{4\pi^2m_q}
|\langle \bar{q} q \rangle_{T,\mu}|^{-1}~.
\end{equation}
The delocalization of the pion wave function due to the melting of the chiral
condensate as expressed in this formula is  {the most important
element of the hadronic freeze-out  mechanism suggested in this work.}

For the nucleon, we shall assume the radius to consist of two
components, a medium independent hard core radius $r_0$ and a pion cloud
contribution
\begin{equation}
r_N^2(T,\mu)=r_0^2+r_\pi^2(T,\mu)~,
\end{equation}
{where from the vacuum values $r_\pi=0.59$ fm and $r_N=0.74$ fm  one gets 
$r_0=0.45$ fm.}

{
For the expansion time scale we adopt a relationship which follows from
entropy conservation, $S=s(T,\mu)~V(\tau_{\rm exp})={\rm const}$,
and a fireball expansion law $V(\tau_{\rm exp})$.
Assuming that
$V(\tau_{\rm exp})\propto \tau^3_{\rm exp}$ one obtains
}
\begin{equation}
\tau_{\rm exp}(T,\mu)=a~s^{-1/3}(T,\mu)~,
\end{equation}
with $a$ being a constant of the order one.

As a  first step,
{we will restrict the discussion to  a medium
consisting of pions and nucleons only, whereby we  apply the above
relationships.}
The generalization to a hadron resonance gas along these
{lines is rather  straightforward.}

{
In the following we introduce properties of the chiral condensate in a
pion-nucleon  and  in a  hadron resonance gas and then applying the freeze-out
condition (\ref{freezeout})
we compare our model results  with  phenomenological findings on hadronic
freeze-out and its conditions.}


\section{Chiral condensate beyond meanfield}

We start from the general definition of the chiral condensate
\begin{eqnarray}
\label{cond}
\langle\bar{q}q \rangle&=&\frac{\partial}{\partial m_0}\Omega(T,\mu)~.
\end{eqnarray}
{The thermodynamical potential  $\Omega(T,\mu)$  can be}
decomposed into contributions from the quark and gluon meanfield and hadronic
(quantized) fluctuations in the meson and baryon channels
\begin{equation}
\label{omega}
\Omega(T,\mu)=\Omega_{\rm MF}(T,\mu)+\Omega_{\rm meson}(T,\mu)
+\Omega_{\rm baryon}(T,\mu)-\Omega(0,0)~.
\end{equation}
The subtraction of $\Omega(0,0)$ removes vacuum divergencies and guarantees
that the thermodynamical potential, i.e., pressure and energy density, of the
vacuum vanish.

The mean field contribution is given by \cite{Hansen:2006ee}
\begin{eqnarray}
\Omega_{\rm MF}(T,\mu) &=&
-  2 N_f \int \frac{d^3p}{(2\pi)^3} \bigg\{3 \varepsilon_p
+  T\; \ln \left[1 + 3 \Phi e^{-\beta (\varepsilon_p-\mu)}
+ 3\Phi e^{-2\beta (\varepsilon_p -\mu)}
+ e^{-3\beta (\varepsilon_p-\mu)} \right]\nonumber \\
&+&   T\; \ln \left[1 + 3 \Phi e^{-\beta (\varepsilon_p+\mu)}
+ 3\Phi e^{-2\beta (\varepsilon_p +\mu)}
+ e^{-3\beta (\varepsilon_p+\mu)} \right]\bigg\}
+ \frac{\sigma^2}{4G} + {\cal{U}}(\Phi ;T) ~,
\label{ommf}
\end{eqnarray}
where the quasiparticle energy is $\varepsilon_p=\sqrt{p^2+m^2(T,\mu)}$,
and $m(T,\mu)=m_0+\sigma(T,\mu)$. In the following we will shorten the
notation by dropping the arguments of $m(T,\mu)=m$ and
$\sigma(T,\mu)=\sigma$.
For the Polyakov-loop potential we take the logarithmic form motivated by the
SU(3) Haar measure,
\begin{equation}
{\cal{U}}(\Phi ,T) = \left[-\,\frac{1}{2}\, a(T)\,\Phi^2 \;+\;b(T)\, \ln(1 -
6\, \Phi^2 + 8\, \Phi^3 - 3\, \Phi^4)\right] T^4 \ ,
\end{equation}
with the corresponding definitions of $a(T)$ and $b(T)$ \cite{Rossner:2007ik}.
In Ref.~\cite{Schaefer:2007pw}, a dependence of the $T_0$ parameter on
the number of active flavors and on the chemical potential has been suggested.
Here we will use $T_0=200$ MeV.
The mean field value of the traced Polyakov loop is then given by
$\bar\Phi = \bar\Phi^\ast = [ 1 + 2\,\cos(\bar\phi_3/T) ]/3$.

The quark mass gap is obtained from the extremum condition
$\partial \Omega/\partial \sigma=0$, equivalent to
\begin{eqnarray}
\label{sigma}
\frac{\sigma}{2G}&=& \frac{6}{\pi^2}
\int dp \, p^2 \frac{m}{\varepsilon_p}
\left[1-f_\Phi^+ -f_\Phi^-\right] ~,
\end{eqnarray}
where the PNJL quark distribution functions are given by
\begin{equation}
\label{f-PNJL}
f^\pm_\Phi=\frac{\Phi[ e^{-\beta (\varepsilon_p-\mu)}
+ 2 e^{-2\beta (\varepsilon_p -\mu)}] + e^{-3\beta (\varepsilon_p-\mu)}}
{1 + 3 \Phi e^{-\beta (\varepsilon_p-\mu)}
+ 3\Phi e^{-2\beta (\varepsilon_p -\mu)}
+ e^{-3\beta (\varepsilon_p-\mu)}}~.
\end{equation}

The values of the Polyakov loop $\Phi$ in the phase diagram are found from a
similar gap equation corresponding to the solution of the extremum condition
$\partial \Omega/\partial \Phi=0$.

The contribution of mesonic fluctuations  has  been discussed in
\cite{Hufner:1994ma,Zhuang:1994dw} for the NJL model and in
\cite{Hell:2008cc,Hansen:2006ee} for the PNJL model.
{Also recently mesonic fluctuations were included in these models within the  functional renormalization group approach \cite{rg1,rg2}.}

{In the following}  we consider only the on-mass-shell meson contributions
 and neglect the continuum correlations beyond the Mott effect.
This is justified because we are actually interested in the
{hadronic freeze-out occurring in a region of the phase diagram which
does not exceed}
the limits of the hadronic phase.
This contribution is given as
\begin{eqnarray}
\label{meson-omega}
\Omega_{\rm meson}(T,\mu)=\sum_{M=\pi,...} d_M
\int \frac{d^3k}{(2\pi)^3} \left\{\frac{E_M(k)}{2}
+  T\; \ln \left[1 - e^{-\beta E_M(k)}\right]\right\}~,
\end{eqnarray}
where the index $M$ denotes the actual meson with degeneracy factor $d_M$.
{We restrict this sum  to the lowest meson state,
the pion, with $E_\pi(k)=\sqrt{k^2+M_\pi^2}$.
}

The contribution of pionic fluctuations to the chiral condensate we obtain from
\begin{equation}
\frac{\partial \Omega_\pi}{\partial m_0}=\langle \bar{q}q\rangle_\pi
=- \frac{M_\pi n_{s,\pi}}{2m_0}+\langle \bar{q}q\rangle_\pi^{\rm vac}~,
\end{equation}
where the GMOR relation has been used to evaluate
$\partial E_\pi(k)/\partial m_0=M_\pi^2/(2m_0 E_\pi(k))$
and the scalar pion density has been defined as
\begin{equation}
  n_{s,\pi}= \frac{d_\pi}{2 \pi^2}\int_0^\infty dp \, p^2\,
\frac{M_\pi}{E_\pi(p)} ~ \frac{1}{e^{\beta E_\pi(p)}-1}~.
\end{equation}
The last term corresponds to a vacuum contribution which gets eliminated
by the vacuum subtraction rule (\ref{omega}) inherent in the definition
of the full condensate (\ref{cond}).

The contribution of baryons to the partition function is considered as an ideal
Fermi gas
\begin{eqnarray}
\label{baryon-omega}
\Omega_{\rm baryon}(T,\mu)=-\sum_{B=N,...} d_B
\int \frac{d^3k}{(2\pi)^3} \left\{\frac{E_B(k)}{2}
+  T\; \ln \left[1 + e^{-\beta (E_B(k)-\mu_B)}\right]\right\}
+  \mu_B \leftrightarrow -\mu_B
.
\end{eqnarray}
For the discussion of the quark condensate we will focus here on the nucleonic
contribution
\begin{equation}
\frac{\partial \Omega_N}{\partial m_0}=\langle \bar{q}q\rangle_N
=-\frac{\sigma_N\, n_{s,N}(T,\mu)}{m_0}+\langle \bar{q}q\rangle_N^{\rm vac}~,
\end{equation}
where we have used the Feynman-Hellman theorem and introduced
the pion-nucleon sigma term,
$\sigma_N=m_0({\partial m_N}/{\partial m_0})=45$ MeV \cite{Cohen:1991nk}.
The nucleon scalar density is given by
\begin{equation}
  n_{s,N}(T,\mu)= \frac{d_N}{2\pi^2}\int_0^\infty dp \, p^2\,\frac{m_N}{E_N(p)}
\left\{f_N(T,\mu)+f_N(T,-\mu) \right\}~,
\end{equation}
where
$f_N(T,\mu)=\{1+\exp[(\sqrt{p^2+m_N^2}-\mu_B)/T]\}^{-1}$
is the nucleon Fermi distribution, $m_N=939$ MeV is the nucleon mass and
$\mu_B=3\mu$ the baryon chemical potential.
The vacuum contribution stemming from the zero-point energy term
gets removed by the vacuum subtraction procedure (\ref{omega}).

{From the above discussion, the chiral condensate in a $\pi -N$ gas can be
obtained as }
\begin{eqnarray}
\label{cond-ultimate2}
-\langle\bar{q}q \rangle&=&\frac{\sigma}{2G}
-\frac{M_{\pi} n_{s,\pi}(T)}{2m_0}-\frac{\sigma_N\, n_{s,N}(T,\mu)}{m_0}~.
\end{eqnarray}
Note,  that using the chiral limit expression for the scalar pion density
$n_{s,\pi}=d_\pi M_\pi T^2/12$, and the GMOR relation (\ref{GMOR}),
we may give this result the form
\begin{eqnarray}
\label{cond-new}
\langle\bar{q}q \rangle&=&
\langle\bar{q}q \rangle_{\rm MF}\left[1-\frac{T^2}{8f^2_\pi(T,\mu)}
-\frac{\sigma_N n_{s,N}(T,\mu)}{M_\pi^2f_\pi^2(T,\mu)}\right]~,
\end{eqnarray}
{which is known from the chiral perturbation theory
\cite{Gasser:1986vb,Buballa:2003qv}, but now with a medium-dependent pion
decay constant.}
Since the modification of the meanfield contribution stemming from the quark
excitations is proportional to $f_\Phi$, it is now effectively suppressed by
the Polyakov-loop, compared to the standard NJL model case.

\begin{figure}[!htpb]
  \centering
  \subfigure[
	]{
    \includegraphics[width=6.5cm]{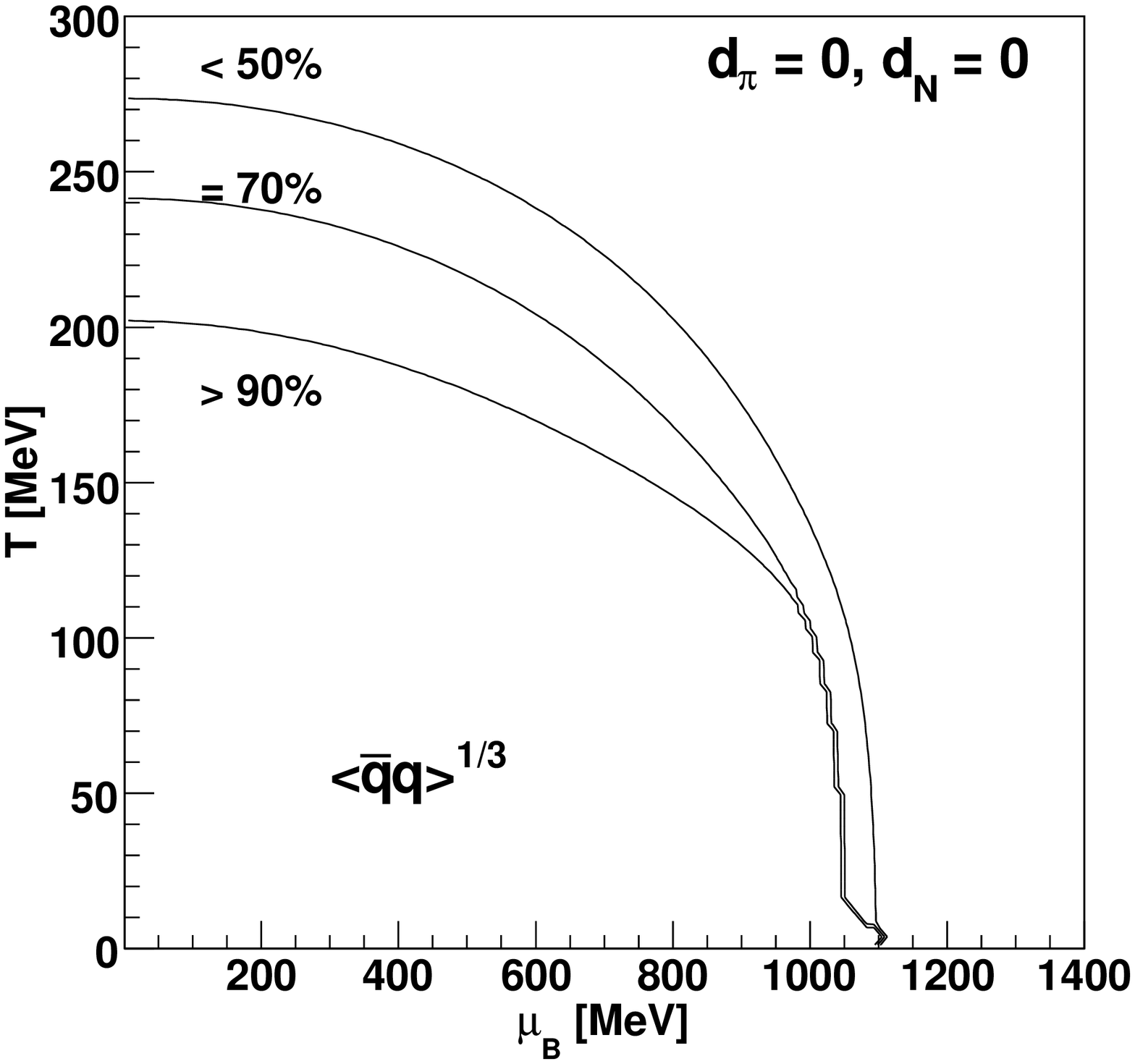}
    \label{fig:PNJL}}
  \subfigure[
	]{
    \includegraphics[width=6.5cm]{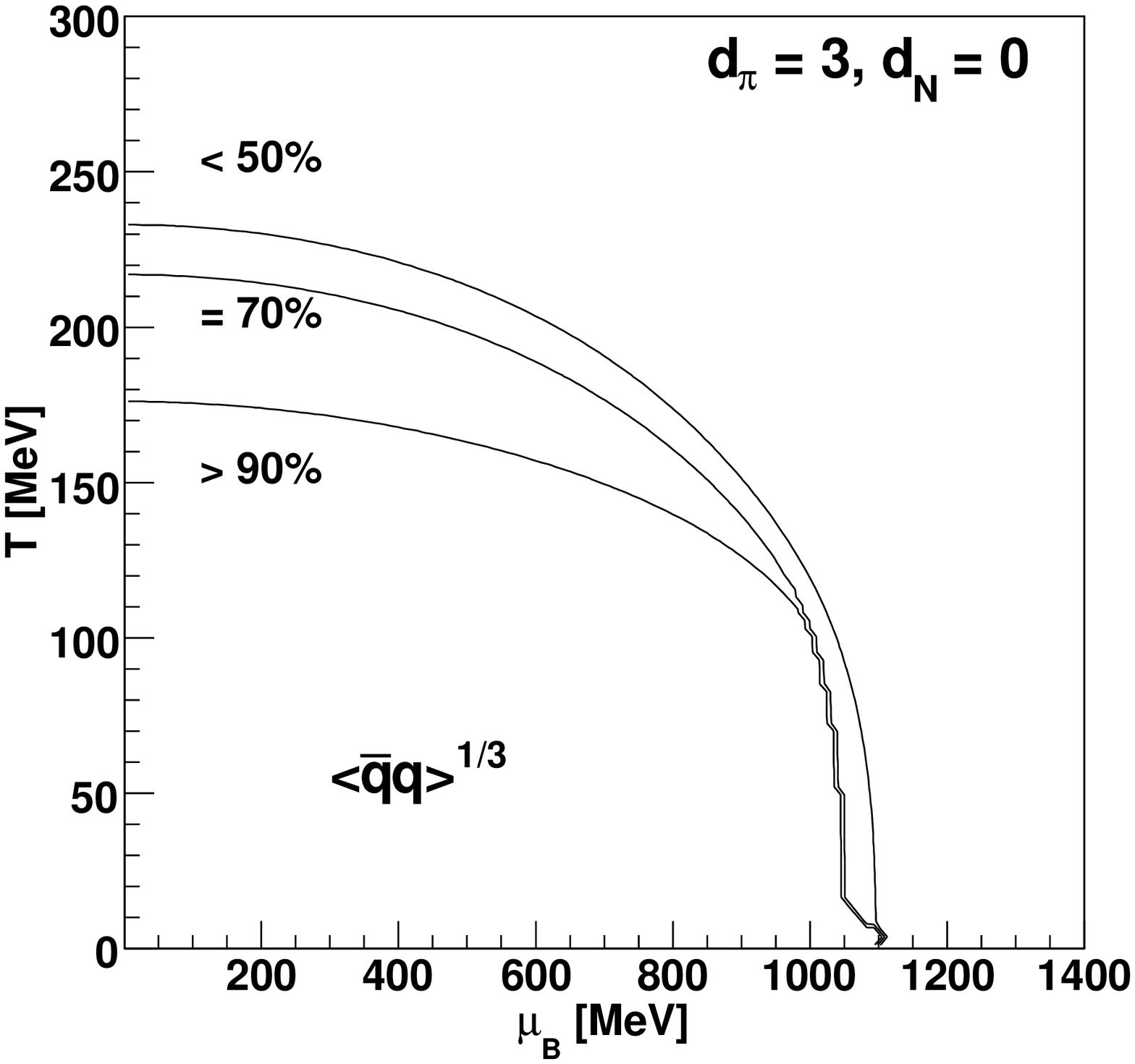}
    \label{fig:PNJL-pi}}
  \subfigure[
	]{
    \includegraphics[width=6.5cm]{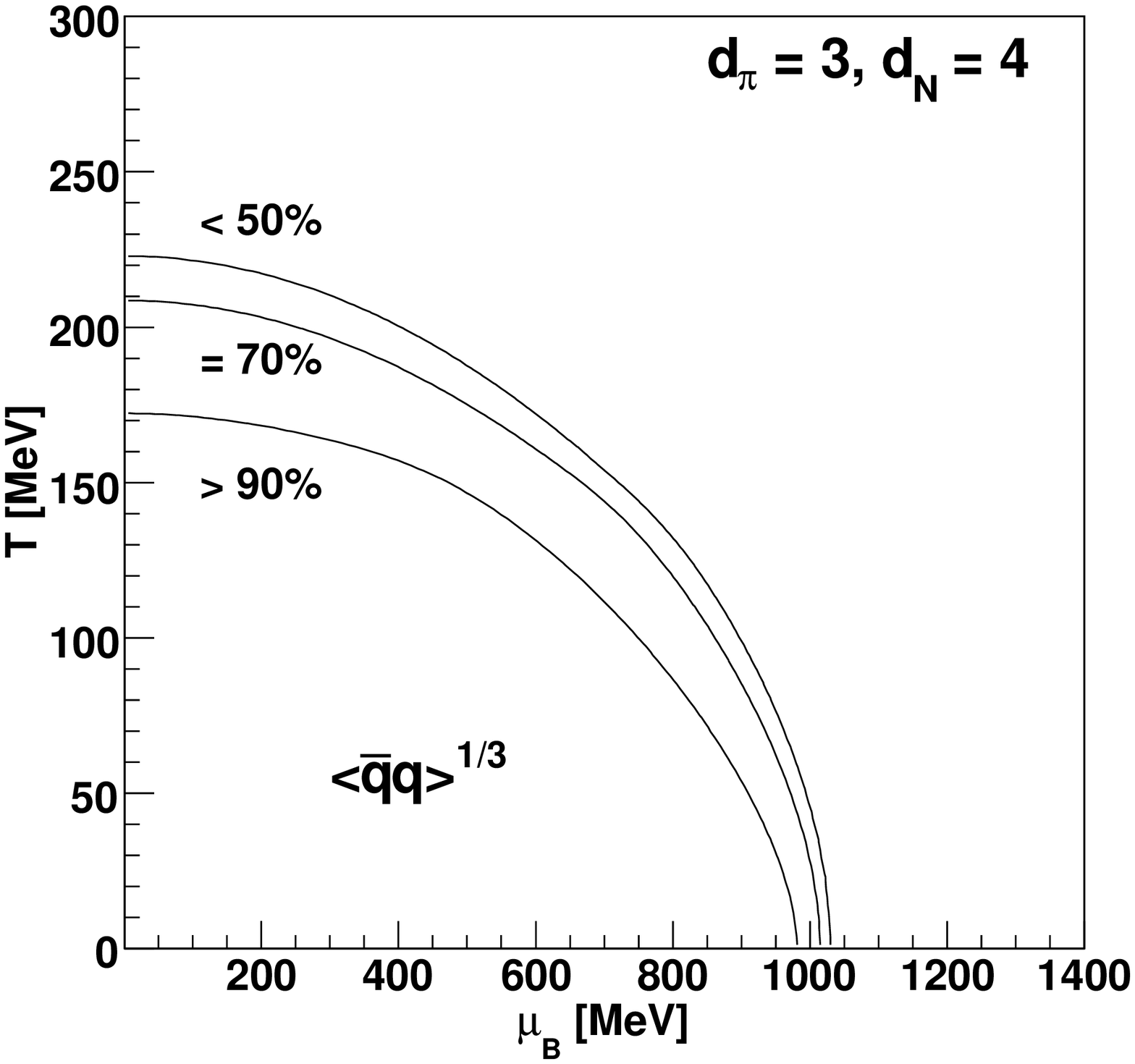}
    \label{fig:PNJL-piN}}
  \subfigure[
	]{
    \includegraphics[width=6.5cm]{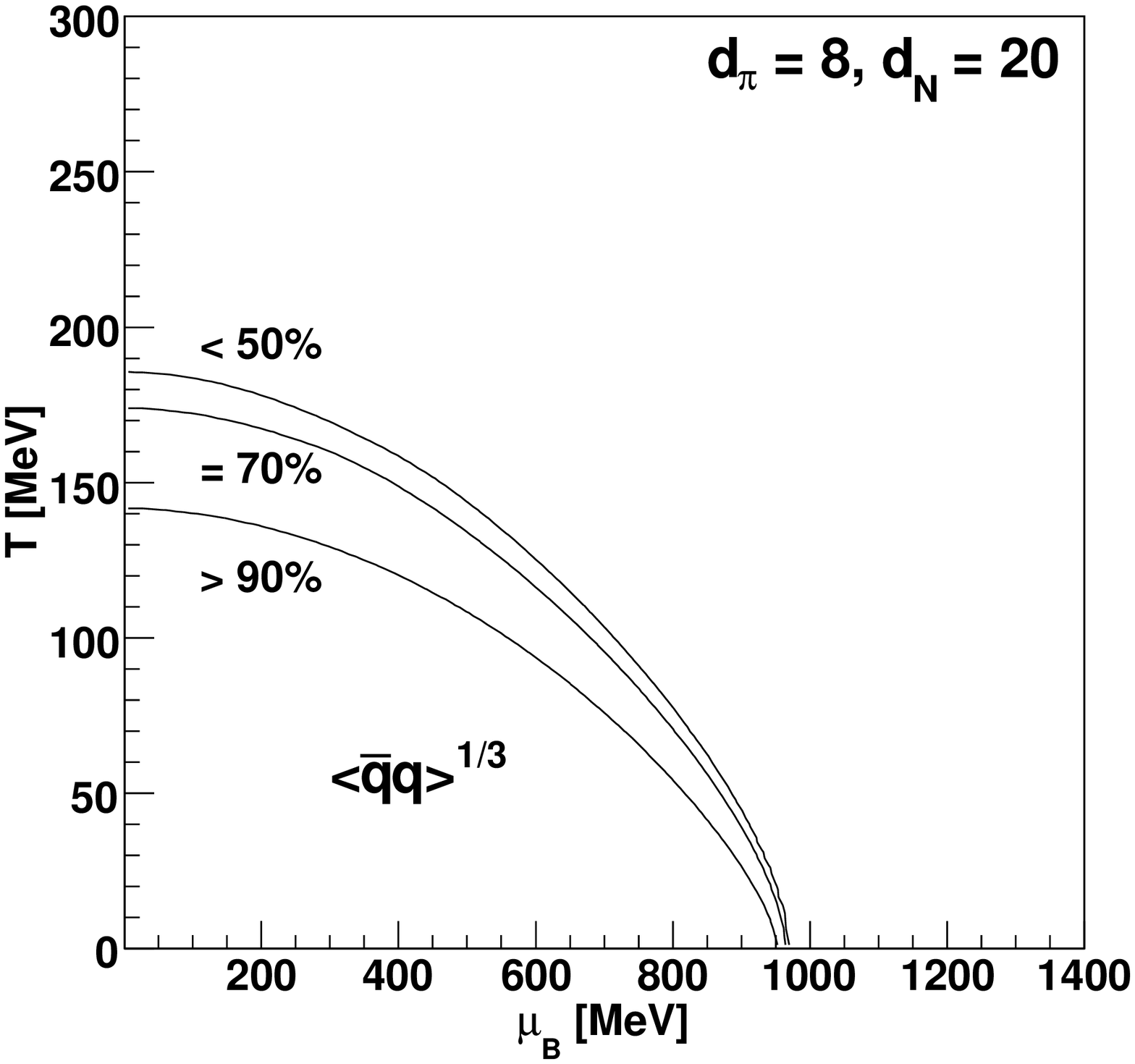}
    \label{fig:PNJL-pi8N20}}
  \caption{Behavior of the chiral condensate $\langle q\bar q \rangle$ in the
PNJL model at meanfield level (a), with pion fluctuations (b), with pion and
nucleon fluctuations (c) and for the schematic resonance gas medium with 8
pionic and 20 nucleonic degrees of freedom (d). 
{Also shown are the corresponding reductions of the chiral condensate relative 
to its vacuum value  for each set of parameters.}}
\end{figure}

Numerical results for the condensate in the $T-\mu$ plane are given in Fig. 1,
where we display contours of equal condensate and indicate its   percentage  reduction relative to the  vacuum value.  We use standard values of the NJL model parametrization, e.g., from
 \cite{Grigorian:2006qe},
with $G\Lambda^2=2.31825$, $\Lambda=602.472$ MeV, $m_0=5.27697$ MeV,
$m_{s,0}=150$ MeV, $M_\pi=140$ MeV, $M_K=495$ MeV, $f_\pi=92.4$ MeV and
$f_K=93.6$ MeV.

\section{The freeze-out curve}

In the following we compare  predictions  of our  model  on the freeze-out conditions with that obtained within  the statistical model analysis of particle yields in heavy ion collisions.

\begin{figure}[!thbp]
    \includegraphics[width=10cm]{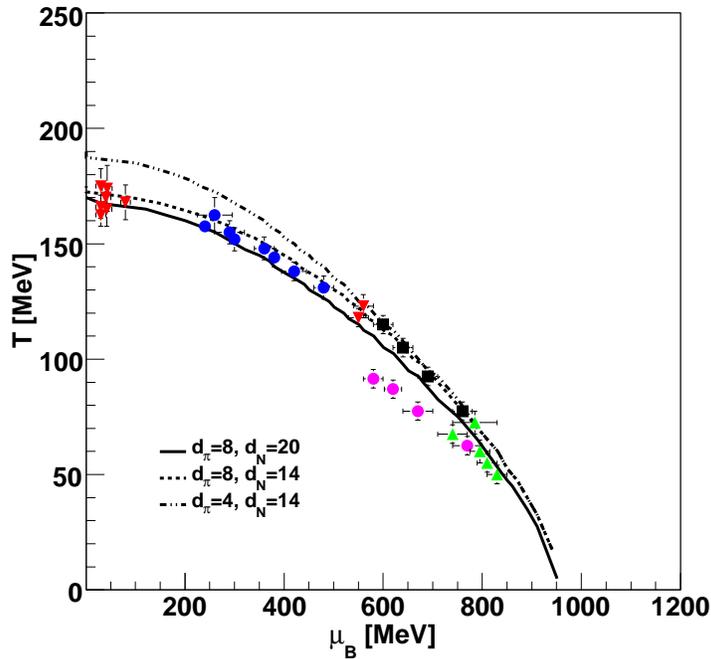}
  \caption{Freeze-out curves according to the kinetic freeze-out
condition, Eq.~(\ref{fo}), for the schematic resonance gas model with
different number of pion ($d_\pi$) and nucleon ($d_N$) degrees of freedom
compared to phenomenological values (symbols) \cite{Andronic:2009gj}.
The lines correspond to different choices for $d_\pi$ and $d_N$ given in the
legend.}
    \label{fig:freezeout}
\end{figure}

{Fig.~\ref{fig:freezeout}  shows  results for  different freeze-out curves
obtained from  the kinetic  condition, Eq.~(\ref{fo}) applied to  the
schematic resonance gas model with variable  numbers of pion ($d_\pi$) and
nucleon ($d_N$) degrees of freedom.
The results are compared with the  phenomenological values.}
From this figure, it is clear that for $d_\pi=8$ and $d_N=14~...~20 $ there is
an excellent correspondence of our freeze-out criterion with phenomenological
values.
Increasing $d_\pi$ results in decreasing   freeze-out temperatures in
the meson-dominated region at small  $\mu_B/T$. On the other hand, increasing
$d_N$ shifts the entire freeze-out curve towards lower $T$ and $\mu_B$.
The freeze-out curves calculated within our model  correspond to a reduction
of the chiral condensate up to
$70\%$ of its vacuum value, c.f. Fig.~1(d), so that the hadron-hadron cross
section in (\ref{freezeout}) according to (\ref{ph}) is roughly twice the
vacuum value.

\begin{figure}[hbtp]
  \centering
  \subfigure[]{
    \includegraphics[width=6.5cm]{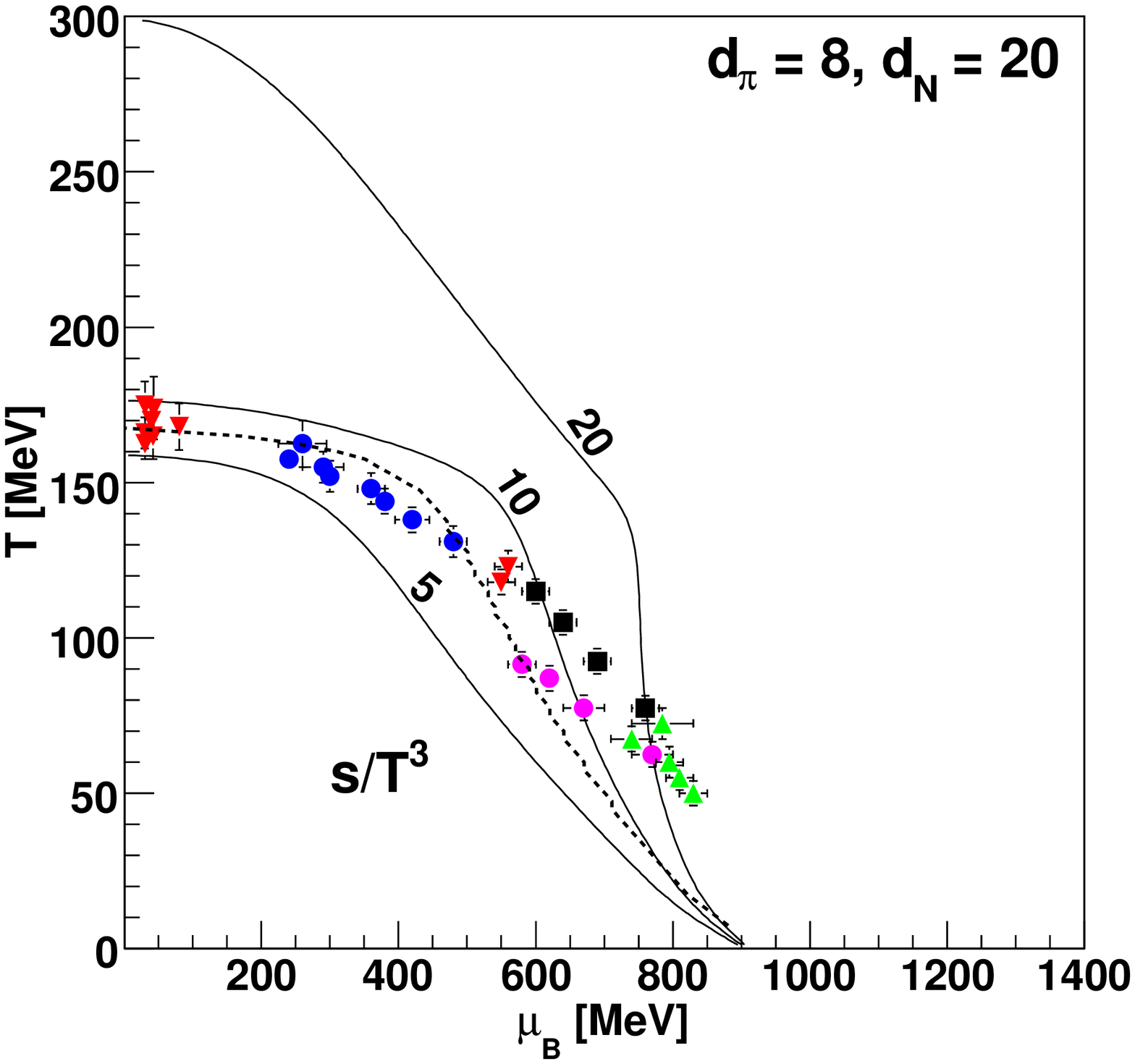}}
\subfigure[]{
    \includegraphics[width=6.5cm]{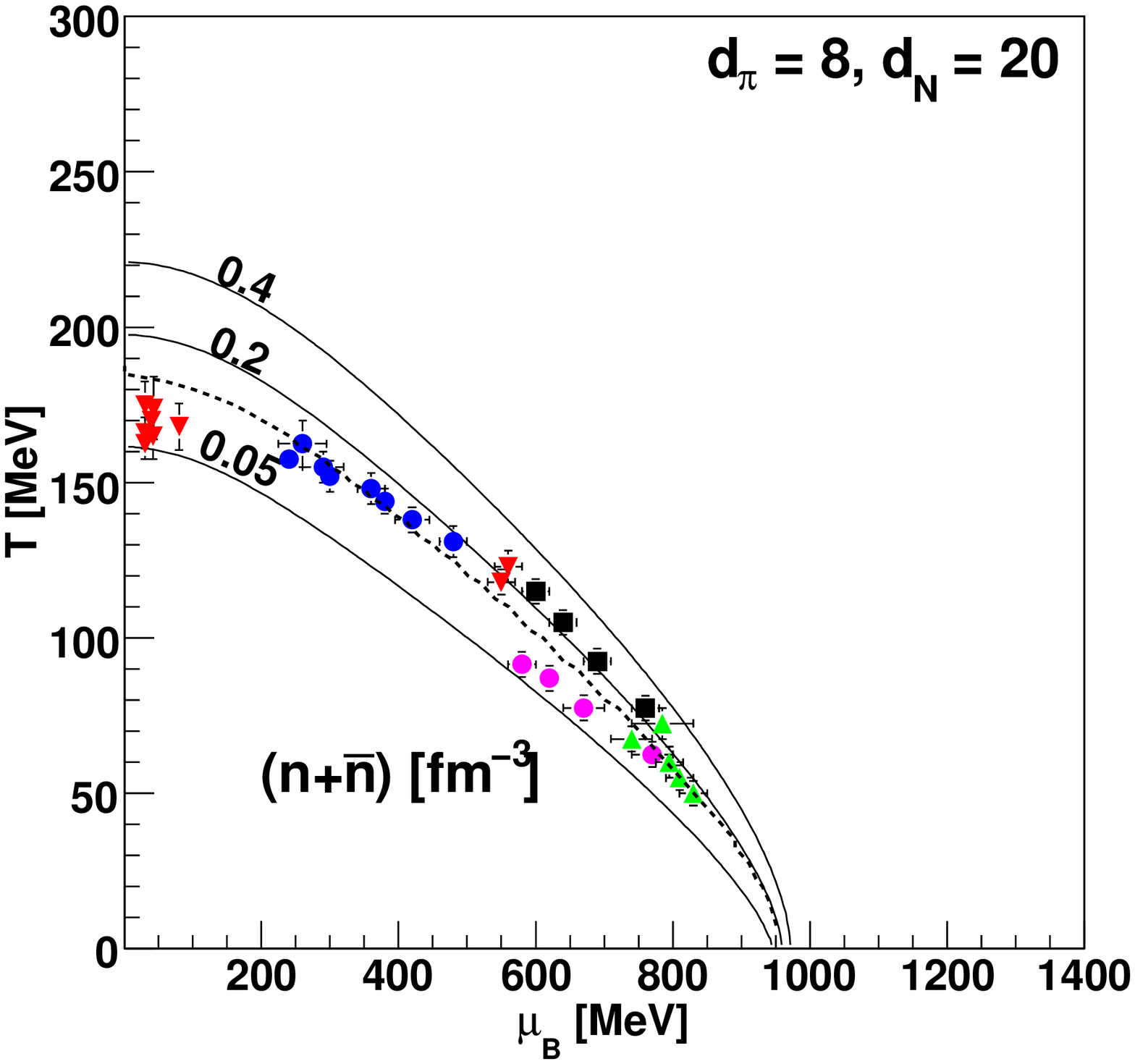}}
  \caption{Entropy $s$ divided by $T^3$ (a) and total baryon density
$n+\bar{n}$ (b) as functions of temperature $T$ and the baryon chemical
potential $\mu_B$ for the schematic resonance gas model with 8 pion and
20 nucleon degrees of freedom.
{The dashed curves represent the phenomenological lines of constant:
$s(T^f,\mu_B^f)/T^3=7$ in (a) and $n_B(T^f,\mu_B^f)+\bar{n}_B(T^f,\mu_B^f)=0.12$ fm$^{-3}$
in (b).}}
    \label{fig:nB-tot}
\end{figure}

{To  verify further our model,} we  evaluate
the entropy density $s(T,\mu)=-\partial \Omega/\partial T$, which according to
the statistical model is related to the freeze-out curve by the
phenomenological condition,  $s(T^{f},\mu_B^{f})/T^3\simeq 7$.
{Fig.~\ref{fig:nB-tot} (left)  shows the comparison of the phenomenological
freeze-out points  with the lines of constant $s(T,\mu)/T^3$ calculated within
our model.
The dashed line in this figure corresponds to  $s(T,\mu)/T^3=7$, which is the
freeze-out line obtained within the hadron resonance gas model.}

Fig.~\ref{fig:nB-tot} (right) shows a similar comparison as the left-hand
figure but assuming freeze-out conditions of constant density of baryons,
$n_B(T^f,\mu_B^f)+\bar n_B(T^f,\mu_B^f)=const.$.
The nucleon density is obtained from
\begin{equation}
n_{N}(T,\mu)=\frac{d_N}{2\pi^2}
\int_0^{\infty}k^2{\rm d}k \frac{1}{1 + e^{\beta (E_N(k)-\mu_B)}}~,
\end{equation}
where $d_N=20$ is the number of nucleonic degrees of freedom in our schematic
resonance gas and
the  antinucleon density, $\bar{n}_N(T,\mu)=n_N(T,-\mu)$.
The result of the calculation with the above parameters is shown in the right
panel of Fig.~\ref{fig:nB-tot}.
From this figure one can  conclude that our effective model reproduces the
phenomenological freeze-out line also when imposing conditions of fixed total
density of baryons.

\section{Conclusion and outlook}

We have developed  chemical freeze-out mechanism which is based on a strong
medium dependence of the rates for inelastic flavor-equilibrating collisions
based on the delocalization of hadronic wave functions and growing
hadronic radii when approaching the chiral restoration.
This approach relates the geometrical (percolation) as well as the
hydrodynamic (softest point) and kinetic (quark exchange) aspects of chemical
freeze-out to the medium dependence of the chiral condensate.
For our model calculations we have employed a beyond-mean-field {\bf effective} extension
of the Polyakov NJL model.
We could demonstrate how the excitation of hadronic resonances initiates the
melting of the chiral condensate which entails a Mott-Anderson type
delocalization of the hadron wavefunctions and a sudden drop in the relaxation
time for flavor equilibration.
{Already for a schematic resonance gas consisting of pions and nucleons with
artificially enhanced numbers of degrees of freedom we could demonstrate that
the kinetic freeze-out condition {can provide} quantitative agreement with the
phenomenological freeze-out curve.
The further development of this approach would clearly require the replacement
of the schematic resonance gas model with the one {based on the full
spectrum of hadronic resonances} and detailed studies of their influence on
the chiral condensate.
{
To this end generalized models for hadron resonance masses are required,
which reveal their dependence on the current quark mass like, e.g., the model
of Leupold \cite{Leupold:2006ih}.
}
However, it is very likely that with such extension the model results
quantifying freeze-out conditions will still be consistent with
phenomenological findings.   }


\subsection*{Acknowledgements}
We acknowledge partial support by the Helmholtz International Center (HIC) for FAIR.
The work of D.B. and K.R. is supported by the Polish Ministry for Science and
Higher Education. 
K.R. received partial support of  ExtreMe Matter Institute (EMMI).

\end{document}